\documentclass[12pt,a4paper]{article}
\usepackage{graphicx}
\usepackage[version=3]{mhchem} 
\usepackage[utf8]{inputenc}   
\usepackage[T1]{fontenc}      
\usepackage[english]{babel}  
                              
\usepackage{a4wide}
\usepackage{latexsym}
\usepackage{epsf}
\usepackage{graphicx}
\usepackage{amsmath,amssymb,amsthm}
\usepackage{amssymb}
\usepackage{braket}
\usepackage{boldline}
\usepackage{cite}
\usepackage{bbm}
\usepackage{hyperref}
\usepackage{verbatim}
\usepackage{xfrac}
\usepackage{tabu}
\usepackage{array,multirow,makecell}
\usepackage{booktabs}
\usepackage{appendix}
\usepackage{setspace}
\setcellgapes{1pt}
\usepackage{ifpdf}
\usepackage{epstopdf}
\makegapedcells
\newcolumntype{R}[1]{>{\raggedleft\arraybackslash }b{#1}}
\newcolumntype{L}[1]{>{\raggedright\arraybackslash }b{#1}}
\newcolumntype{C}[1]{>{\centering\arraybackslash }b{#1}}

\newenvironment{changemargin}[2]{\begin{list}{}{%
\setlength{\topsep}{0pt}%
\setlength{\leftmargin}{0pt}%
\setlength{\rightmargin}{0pt}%
\setlength{\listparindent}{\parindent}%
\setlength{\itemindent}{\parindent}%
\setlength{\parsep}{0pt plus 1pt}%
\addtolength{\leftmargin}{#1}%
\addtolength{\rightmargin}{#2}%
}\item }{\end{list}}
\newcommand{\subf}[2]{%
  {\small\begin{tabular}[t]{@{}c@{}}
  #1\\#2
  \end{tabular}}%
}

\newcommand\pubdate{\rightline{IPhT-T17/028}}

\def\saclay{Institut de Physique Théorique, Université Paris Saclay, CEA \\
Orme des Merisiers, F-91191 Gif-sur-Yvette, France \\\medskip Département de Physique, \'Ecole Normale Sup\'erieure de Lyon, Universit\'e de Lyon\\ 46 All\'ee d'Italie, F-69364 Lyon, cedex 07, France}

\def\Title#1{\begin{center} {\begin{spacing}{1.5}\Large #1 \end{spacing}} \end{center}}
\def\Author#1{\begin{center}{ \textbf{ #1}} \end{center}}
\def\Address#1{\begin{center}{ \it #1} \end{center}}

\newenvironment{Abstract}{\begin{quotation} \begin{center} 
             \textbf{Abstract}\end{center}\bigskip 
      }{ \end{quotation}}

\begin{document}
\begin{titlepage}
\pubdate

\vfill
\Title{\textbf{Four-center bubbled BPS solutions with a Gibbons-Hawking base}}

\Author{Pierre Heidmann}
\Address{\saclay \\ 
\vspace{0.5cm}
\begin{center}\normalfont{pierre.heidmann@cea.fr}\end{center}}
\vfill
\begin{Abstract}
We construct four-center bubbled BPS solutions with a Gibbons-Hawking base space. We give a systematic procedure to build scaling solutions: starting from three-supertube configurations and using generalized spectral flows and gauge transformations to extend to solutions with four Gibbons-Hawking centers. This allows us to construct very large families of smooth horizonless solutions that have the same charges and angular momentum as supersymmetric black holes with a macroscopically large horizon area. Our construction reveals that all scaling solutions with four Gibbons Hawking centers have an angular momentum at around 99\% of the cosmic censorship bound. We give both an analytical and a numerical explanation for this unexpected feature.
\end{Abstract}

\vfill
\end{titlepage}

\tableofcontents

\section{Introduction}
Multi-center BPS bubbling solutions brought about important breakthroughs in the understanding of the microstate geometries of black holes. In the context of the fuzzball proposal (see \cite{Mathur:2005zp, Bena:2007kg,Mathur:2008nj,Balasubramanian:2008da} for reviews), they allow to understand the quantum structure of black holes by resolving singularities of classical black hole solutions into smooth, horizonless geometries. The main features of the geometries which replace the singularity are topologically non-trivial cycles called ``bubbles'' maintained by fluxes.\\
The physics of bubbled geometries is highly constrained by certain regularity conditions, known as the bubble equations, and the ``no-CTC condition'' \cite{Bena:2005va,Denef:2000nb} which make the charges, the angular momenta and the positions of the centers all interdependent. Furthermore, to obtain horizonless microstate solutions with a black-hole-like throat, the bubbles need to ``scale'', that is to say, to shrink in the \(\mathbb{R}^3\) base of the Gibbons-Hawking (GH) space \cite{Bena:2006kb,Bena:2007qc}. Consequently, it is important to investigate the relation between charges, angular momenta and positions to understand the miscrostate geometries.\\
Bubbling BPS solutions with a Gibbons-Hawking base in five dimensions have been meticulously studied since a few years. A very large class of microstate geometries of supersymmetric black holes have been constructed from those solutions. The solutions can be written in a M2-M2-M2 frame or in other duality frames such as D1-D5-P or as D0-D4-F1 (for a review see \cite{Bena:2007kg}).\\ 
Finding specific examples of smooth horizonless multi-bubble solutions that have the same charges as a large black hole is quite non-trivial. Indeed, most solutions one can construct by putting fluxes on a multi-center Gibbons-Hawking base have angular momenta larger than the black hole cosmic censorship bound and hence can not be thought of as black hole microstate geometries. Trying to modify these fluxes to reduce the angular momenta gives very often rise to closed timelike curves. This is why in the literature, one can only find sparse specific examples of multi-center BPS solutions \cite{Bena:2006kb,Bena:2007kg,Vasilakis:2011ki,Bianchi:2017bxl}.  For example, one seven-center BPS three-charge black hole solution was built in \cite{Bena:2006kb}. The solution is scaling and the angular momentum \(J\) is about \(84\%\) of its maximal value. However, some distances between the centers are 1000 times bigger than others. \\

In this article, we present a systematic construction and analysis of the largest known family of scaling four-center smooth horizonless solutions that have the same charges as large black holes. Our construction allows us to easily build scaling four-center BPS solutions with any aspect ratios between the centers. We study the charges and angular momentum of the solutions, as well as the entropy parameter, $\mathcal{H}$, given by
\begin{equation}
\mathcal{H} \: \equiv\: \frac{Q_1 Q_2 Q_3 \:-\: J^2}{Q_1 Q_2 Q_3}.
\end{equation}
The entropy parameter $\mathcal{H}$ indicates how far is the solution from its corresponding maximally spinning black hole solution. Our first surprising result is that all our solutions have a fine-tuned entropy parameter around 0.01. Several arguments lead us to the idea that four Gibbons-Hawking center BPS solutions which do not have an entropy parameter close to 0 are really rare. Moreover, we will show that when there is no difference in scales between the inter-center distances, all BPS solutions with four GH centers have an entropy parameter around 0,01. \\
The main advantage of BPS solutions with four GH centers is that they can be related by spectral flows to three supertubes in Taub-NUT (TN). Indeed, three-supertube constructions in Taub-NUT are easier to analyze than general four-GH center solutions and spectral flows are specific transformations which ensure that CTC are not introduced. Consequently, one can mostly concentrate on three-supertube constructions. \\
In Section two, we construct several classes of specific solutions with three supertubes in TN. We give a systematic method to build examples of such scaling solutions. In Section three, we use generalized spectral flow to construct bubbled BPS solutions with four GH centers from the three-supertube solutions. We then illustrate our method by constructing several classes of solutions with four GH centers. These solutions have an angular momentum 99\% of the cosmic censorship bound. In Section four, we present a numerical and an analytical proof that all solutions with four GH centers and black hole charges have such large angular momenta.


\section{Particular solutions with three BPS supertubes in Taub-NUT}

\subsection{Three-supertube BPS solutions in Taub-NUT}

The notations and the results are taken from \cite{Bena:2009en,Vasilakis:2011ki}. We will recall the three-charge solutions for a system of three supertubes in Taub-NUT. In M-theory, the three charges correspond to three M2 branes wrapping three orthogonal 2-tori inside a 6-torus. The eleven-dimensional metric is:

\begin{equation}
ds_{11}^2 \:=\: -\left(Z_1 Z_2 Z_3 \right)^{-2/3} \, \left( dt \:+\: k \right) ^2 \:+\: \left(Z_1 Z_2 Z_3 \right)^{1/3} \,ds_4^2 \:+\: \left(Z_1 Z_2 Z_3 \right)^{1/3} \,\sum_{I=1}^3 \frac{dx_{I}^2 \:+\: dy_{I}^2}{Z_I},
\label{11dmetric}
\end{equation}
where \(dx_{I}^2 \,+\, dy_{I}^2\) is the metric of the I$^{th}$ 2-Torus and \(ds_{4}^2\) is the four-dimensional Gibbons-Hawking metric with one Gibbons-Hawking center:
\begin{equation}
ds_4^2 \:=\: V^{-1} \! \left( d\psi \:+\: A \right) ^2 \:+\: V \left( dx^2 \:+\: dy^2 \:+\: dz^2 \right)
\end{equation}
with
\begin{equation}
V \:=\: h \:+\: \frac{q}{r}, \qquad \vec{\nabla} \times \vec{A} \:=\: \vec{\nabla}V 
\end{equation}
The solutions have a three-form potential \(\mathcal{A}\):
\begin{equation}
\mathcal{A} \:=\: \sum_{I=1}^3 A^{(I)} \:\wedge\: dx_{I}^2 \:\wedge\: dy_{I}^2.
\end{equation}
The one forms $k$ and \(A^{(I)}\) depend only on the four space coordinates transverse to the 6-Torus. The dipole field strengths \(\Theta^{(I)}\) are defined as:
\begin{equation}
\Theta^{(I)} \:\equiv\: d A^{(I)} \:+\: d\left(\frac{dt \:+\: k }{Z_I}\right).
\end{equation}
In order to have BPS solutions, the fields must satisfy the following equations \cite{Gutowski:2004yv,Bena:2004de},
\begin{equation}
\begin{aligned}
\Theta^{(I)} \:&=\: \star_4 \Theta^{(I)} \\
\hat{\nabla}^2 Z_I \:&=\: \frac{1}{2} C_{IJK} \star_4 \left(\Theta^{(J)} \wedge \Theta^{(K)} \right) \\
dk \:+\: \star_4 dk \:&=\: Z_I \Theta^{(I)}.
\label{eqofmotionfirst}
\end{aligned}
\end{equation}
The Hodge dual \(\star_4\) is the Hodge dual of the four-dimensional Gibbons-Hawking space, and \(C_{IJK}\) is the absolute value of the constant symmetric tensor. \\
We consider axisymmetric supertube configurations. The positions of the supertube centers are given by the distances \(a_1\), \(a_2\) and \(a_3\) on the z-axis in the following order
\begin{equation}
a_1 \:>\: a_2 \:>\: a_3 \:>\: 0.
\end{equation}

The BPS equations \eqref{eqofmotionfirst} allow to consider three types of two-charge supertubes. Each supertube carries a dipole charge \(k_I\) and two electric charges \(Q_j^{(I)}\) at the points $j \neq I$. Consequently, the dipole strengths \(\Theta^{(I)}\) and the scalar fields \(Z_I\) are sourced by harmonic functions which we call \(K^I\) and \(L_I\)

\begin{equation}
\begin{split}
& K^{I} \:=\: \alpha_I \:+\: \frac{k_I}{r_I}. \\
& L^{1} \:=\: 1 \:+\: \frac{Q_2^{(1)}}{4 r_2} \:+\: \frac{Q_3^{(1)}}{4 r_3} \quad,\qquad \textrm{\(L^2\) and \(L^3\) by permutation,}
\label{K}
\end{split}
\end{equation}
where \(r_I\) is the distance to the I\(^{th}\) center
\begin{equation}
r_I \:\equiv\: \sqrt{x^2 \!+\! y^2 \!+\! \left(z-a_I\right)^2}.
\end{equation}
The complete solution for the warp factors is:

\begin{equation}
\begin{split}
 Z_I &\:=\: L_I \:+\: \frac{1}{2}C_{IJK} \frac{K^J K^K}{V} \\
     &\:=\: 1 \:+\: \sum\limits_{J \ne I} \frac{Q_J^{(I)}}{4 r_J} \:+\: \frac{C_{IJK}}{h\!+\! \frac{q}{r}}\left( \frac{k_J k_K}{2 r_J r_K} \:+\: \frac{\alpha_J \alpha_K }{2} \:+\: \frac{\alpha_J k_K}{r_K}\right),
\label{Zexpression}
\end{split}
\end{equation}
and 
\begin{equation}
k \:=\: \mu \left( d\psi \!+\! A \right) \:+\: \omega,
\label{kexpression}
\end{equation}
with 
\begin{equation}
\mu \:=\: \frac{1}{6} V^{-2} C_{IJK} K^I K^J K^K \:+\: \frac{1}{2} V^{-1} K^I L_I \:+\: M,
\label{muexpression}
\end{equation}
where M is another harmonic function which we will take to be 
\begin{equation}
M \:=\: m_\infty \:+\: \frac{m_0}{r} \:+\: \sum\limits_{j=1}^3 \frac{m_j}{r_j},
\end{equation}
and \(\omega\) is another vector which satisfies the following equation in the \(\mathbb{R}^3\) base of TN:

\begin{equation}
\vec{\bigtriangledown} \times \omega \:=\: V \vec{\bigtriangledown}\mu \:-\: \mu\vec{\bigtriangledown}V - V\sum\limits_{I=1}^3 Z_I \vec{\bigtriangledown} \left( \frac{K^I}{V} \right).
\label{omegaexpression}
\end{equation}

Thus, $\mu$ is
\begin{equation}
\begin{split}
 \mu \:=\: &\frac{1}{V^2} \left( \frac{k_1 k_2 k_3}{r_1 r_2 r_3} \:+\: \alpha_1 \alpha_2 \alpha_3 \:+\: \frac{C_{IJK}}{2}\left( \frac{\alpha_I k_J k_K }{r_J r_K} \:+\: \frac{\alpha_I \alpha_J k_K }{r_K}\right)\right) \\
& \:+\: \frac{C_{IJK}}{4V}\left(\alpha_I \:+\: \frac{k_I}{r_I}\right)\left(1 \:+\: \frac{Q_J^{(I)}}{4 r_J} \:+\: \frac{Q_K^{(I)}}{4 r_K} \right)\:+\: M.   
\end{split}
\end{equation}


\subsection{Bubble equations and regularity conditions}

In addition to the BPS equations \eqref{eqofmotionfirst}, the solutions must satisfy several regularity conditions that we recall in this section. First, divergent terms along the \(\psi\)-fiber in the metric must vanish. Second, closed timelike curves (CTC) must be absent.  We organize the harmonic functions in a symplectic vector H:
\begin{equation}
\begin{split}
H \:&=\: \left( H^0 , H^I , H_I , H_0 \right) \:\equiv\:  \left( V , K^I , L_I , 2M \right) \\
&=\: \hat{h} \:+\: \sum\limits_{j=0}^3 \frac{\Gamma_j}{r_j},
\end{split}
\end{equation}
with 
\begin{equation}
\hat{h} \:=\: \left( h,\alpha_1,\alpha_2,\alpha_3;1,1,1,2 m_\infty \right), \quad \Gamma_0 \:=\: \left( q,0,0,0;0,0,0,2 m_0 \right), \quad \Gamma_j \:=\: \left(0,k_j;\frac{1}{4}Q_j,2 m_j\right).
\end{equation}
The conditions of regularity at the singularities \(r_j \!\rightarrow\: 0\) \(j=\left\{0,1,2,3\right\}\) with \(r_0 \!=\! r\) are summed up by the following equations \cite{Denef:2000nb}:
\begin{equation}
\sum\limits_{J=0}^3 \frac{\langle \Gamma_I , \Gamma_J \rangle}{r_{IJ}} \:=\: -\langle \Gamma_I , \hat{h} \rangle  \qquad I \!=\! \left\{0,1,2,3\right\},
\label{BuEQ}
\end{equation}
where \(\langle\:,\:\rangle\) is a symplectic product defined as follows
\begin{equation}
    \langle A,B \rangle \:=\:  A_0 B_8 - A_8 B_0 +A_I B_{8-I} - A_{8-I} B_I. 
\end{equation} \\
Using the notation
\begin{equation}
\Gamma_{IJ} \:=\: k_I Q_J^{(I)} \:-\: k_J Q_I^{(J)},
\end{equation}
we obtain the following bubble equations
\begin{equation}
\begin{split}
& \frac{\Gamma_{12}}{r_{12}} \:+\: \frac{\Gamma_{13}}{r_{13}} \:=\: 8 m_1 \left( h \!+\! \frac{q}{a_1}\right) \:-\: 4 k_1 \:+\: Q_1^{(2)} \alpha_2 \:+\: Q_1^{(3)} \alpha_3\\
& \frac{\Gamma_{21}}{r_{12}} \:+\: \frac{\Gamma_{23}}{r_{23}} \:=\: 8 m_2 \left( h \!+\! \frac{q}{a_2}\right) \:-\: 4 k_2 \:+\: Q_2^{(1)} \alpha_1 \:+\: Q_2^{(3)} \alpha_3\\
& \frac{\Gamma_{32}}{r_{23}} \:+\: \frac{\Gamma_{31}}{r_{13}} \:=\: 8 m_3 \left( h \!+\! \frac{q}{a_3}\right) \:-\: 4 k_3 \:+\: Q_3^{(1)} \alpha_1 \:+\: Q_3^{(2)} \alpha_2\\
& q\left( m_\infty \:+\: \sum\limits_{j=1}^3 \frac{m_j}{a_j} \right) \:=\: m_0 h.
\end{split}
\label{BubbleEquation}
\end{equation} 

Furthermore, all the parameters in \(M\) are constrained to avoid Dirac strings at each pole and to avoid divergences along the \(\psi\)-fiber \cite{Bena:2009en}:
\begin{equation}
\begin{split}
& m_1 \:=\: \frac{Q_1^{(2)}Q_1^{(3)}}{32 k_1},\qquad m_2 \:=\: \frac{Q_2^{(1)}Q_2^{(3)}}{32 k_2}, \qquad m_3 \:=\: \frac{Q_3^{(2)}Q_3^{(1)}}{32 k_3}, \\
& m_0 \:=\: 0, \qquad m_\infty \:=\: - \sum\limits_{j=1}^3 \frac{m_j}{a_j}.
\end{split}
\label{MConstants}
\end{equation}

Finally, the absence of CTCs is satisfied when the quartic invariant $I_4$ is greater than $\omega^2$ in the three-dimensional base space of the solution \cite{Berglund:2005vb}:
\begin{equation}
I_4 \:\equiv\: Z_1 Z_2 Z_3 V \:-\: \mu^2 V^2 \:>\: \omega^2.
\label{CTCcondition}
\end{equation}


\subsection{Examples of solutions}
The goal of this section is to construct several examples of supertube charges and dipole charges such that the bubble equations have physical solutions. In particular, we will focus on solutions with no difference in scales between the inter-center distances. We know from \cite{Bena:2006kb} that the depth of the solution's throat is inversely related to the distance between the centers in $\mathbb{R}^3$. Thus, the centers must cluster to form a concentrated object in order for the microstate geometry to look like a black hole when seen from large distances. Several methods exist for constructing such solutions. In general solutions scale when the centers move away from axisymmetry. However, to scale axisymmetric solutions we will slightly change the value of the charges, as in \cite{Bena:2006kb}.


\subsubsection{Specific choice of parameters and ``scaling conditions''}
\label{subsubsec:specificchoice}
We consider the following choice of charges and dipole charges inspired from the choice made in \cite{Vasilakis:2011ki} 

\begin{equation}
\begin{split}
& k_1 \:=\: k_3 \:=\: -\:k_2 \:=\: k \:>\: 0 \\
& Q_1^{(2)} \:=\: Q_3^{(2)} \:=\: \alpha \:>\: 0 \\ 
& Q_3^{(1)} \:=\: Q_2^{(3)} \:=\: \beta \:>\: 0 \\
& Q_2^{(1)} \:=\: Q_1^{(3)} \:=\: \gamma \:>\: 0.
\end{split}
\label{SystemOfChargesBPS}
\end{equation} 

Finding an explicit scaling condition is crucial. Indeed, one can find many solutions of the bubble equations which do not scale. However, one must be able to go to a scaling limit to build a solution which looks like a black hole from far away. \\
For this purpose, we solve the bubble equations \eqref{BubbleEquation} by computing the charges for a given cluster of supertubes infinitely close to the Gibbons Hawking center, that is to say in the limit \(a_i \!\rightarrow\: 0\). So we consider an arbitrary arrangement of distances
\begin{equation}
a_j \:=\: \lambda \:d_j \qquad j\!=\! \left\{1,2,3\right\},
\end{equation} 
with
\begin{equation}
\lambda \:\ll\: 1 \: , \qquad d_1 \:>\: d_2 \:>\: d_3.
\end{equation}
We linearize the bubble equation \eqref{BubbleEquation} in powers of \(\lambda\)
\begin{equation}
\begin{split}
& \frac{1}{\lambda} \left( \frac{\alpha +\gamma}{d_1-d_2} \:+\: \frac{\beta - \gamma}{d_1-d_3} \:-\: \frac{q}{4 k^2 d_1} \alpha \gamma \right) \:=\: \frac{h}{4 k^2} \alpha \gamma \:-\: 4 \:+\: \frac{\alpha\alpha_2 + \gamma\alpha_3}{k}\\
& \frac{1}{\lambda} \left( - \frac{\alpha +\gamma}{d_1-d_2} \:-\: \frac{\alpha + \beta}{d_2-d_3} \:+\: \frac{q}{4 k^2 d_2} \beta \gamma \right) \:=\: - \frac{h}{4 k^2} \alpha \gamma \:+\: 4\:+\: \frac{\gamma\alpha_1 + \beta\alpha_3}{k}\\
& \frac{1}{\lambda} \left( \frac{\gamma - \beta}{d_1-d_3} \:+\: \frac{\alpha + \beta}{d_2-d_3} \:-\: \frac{q}{4 k^2 d_3} \alpha \beta \right) \:=\: \frac{h}{4 k^2} \beta^2 \:-\: 4 \:+\: \frac{\beta\alpha_1 + \alpha\alpha_3}{k}.\\
\label{BubbleEquation2}
\end{split}
\end{equation} 
Consequently, the left-hand sides of the previous equations must be close to 0 if we want \(\lambda\) to go to 0. With the specific choice of charges and dipole charges \eqref{SystemOfChargesBPS}, analytic solutions can be found

\begin{equation}
\begin{split}
& \alpha \:=\: \frac{4 k^2}{q}\frac{a_2^{\;3}a_3 \:+\: a_1^{\;2}a_3\left(2a_2 - a_3\right) \:+\: a_1 a_2 \left( a_2^{\;2} - 5a_2 a_3 + 2 a_3^{\;2}    \right)}{a_2 \left(a_1 -a_2 \right) \left(a_1 -a_3 \right) \left(a_2 -a_3 \right)} \:+\: \mathcal{O}\left(\lambda\right)\\
& \beta\:=\:\frac{4 k^2}{q}\frac{a_2^{\;3}a_3 \:+\: a_1^{\;2}a_3\left(2a_2 - a_3\right) \:+\: a_1 a_2 \left( a_2^{\;2} - 5a_2 a_3 + 2 a_3^{\;2}    \right)}{a_1 \left(a_1 -a_2 \right) \left(a_2 -a_3 \right)^2} \:+\: \mathcal{O}\left(\lambda\right)\\
& \gamma \:=\: \frac{4 k^2}{q}\frac{a_2^{\;3}a_3 \:+\: a_1^{\;2}a_3\left(2a_2 - a_3\right) \:+\: a_1 a_2 \left( a_2^{\;2} - 5a_2 a_3 + 2 a_3^{\;2}    \right)}{a_3 \left(a_1 -a_2 \right)^2 \left(a_2 -a_3 \right)} \:+\: \mathcal{O}\left(\lambda\right).
\end{split}
\label{constraintscaling}
\end{equation} 

The main difference between our solutions and the solutions in \cite{Vasilakis:2011ki} is that the three centers and the Gibbons-Hawking center are scaling together in our solutions, while in \cite{Vasilakis:2011ki} only supertubes were scaling. This will prove to be valuable when we construct scaling solutions with four GH centers. Those formulas are useful to have an idea about the values of charges we need to choose to obtain a particular scaling solution with three supertubes in Taub-NUT. We apply those formulas in the next sections to find specific examples of solutions with no difference in scales between the inter-center distances.


\subsubsection{Solutions with $h \,=\, 0$ and with no constant terms in all $K^I$}

Imposing \(h \,=\, 0\) and \(q \,=\, 1\) is principally motivated by having a Gibbons-Hawking metric which looks like \(\mathbb{R}^4\) at infinity. The first example of charges and dipole charges is
\begin{equation}
\begin{split}
& h \:=\: 0 \\
& q \:=\: 1 \\
& k_1 \:=\: k_3 \:=\: -\:k_2 \:=\: k \:=\: 100\\
& Q_1^{(2)} \:=\: Q_3^{(2)} \:=\: \alpha \:=\: 70\,000  \\ 
& Q_3^{(1)} \:=\: Q_2^{(3)} \:=\: \beta \:=\: 100\,000  \\
& Q_2^{(1)} \:=\: Q_1^{(3)} \:=\: \gamma \:=\: 233\,333.
\end{split}
\label{SystemOfCharges}
\end{equation} 

We solve numerically the bubble equations \eqref{BubbleEquation}:
\begin{equation}
a_1 \:=\: 8.46360\ldots\times10^{-2} \,, \qquad a_2 \:=\: 4.02091\ldots\times10^{-2}\,, \qquad a_3 \:=\: 1.80731\ldots\times10^{-2}.
\end{equation}
One can check straightforwardly by plotting the quartic invariant \(I_4\) and $\omega^2$ in the \(\mathbb{R}^3\) space that the no-CTC condition \eqref{CTCcondition} is satisfied. \\
We notice that many other different scaling solutions can be built from \eqref{CTCcondition}.
\begin{itemize}
    \item \textbf{Charges, angular momentum and entropy of the solution}
\end{itemize}

We give also the charges and the angular momentum of the solution:

\begin{equation}
 \begin{split}
     & Q_1 \:=\: Q_3 \:=\: \frac{293\;333}{4}\\
     & Q_2 \:=\: 45\;000 \\
     & J \:=\: \frac{44\;666\;625}{8}.\\
 \end{split}
\end{equation}

The Bekenstein-Hawking entropy of the corresponding black hole is then 

\begin{equation}
 S \:=\: 2\pi \sqrt{Q_1 Q_2 Q_3 \;-\; J^2} \:=\: 9.12309\ldots\times10^{7}.
\end{equation}

We compute the entropy parameter \(\mathcal{H}\) which informs how far the charges of the solution are from those of the corresponding maximally-spinning black hole.

\begin{equation}
 \mathcal{H} \:=\: 0.87\ldots.
\end{equation}

\begin{itemize}
    \item \textbf{Scaling solutions}
\end{itemize}

Finally, we scale the solution. We fine-tune the values of the initial charges, we solve the bubble equation, and then we check the absence of CTCs. The scaling process is summed up in the following table:
\bigbreak
\begin{changemargin}{-1cm}{-1cm}
\begin{tabular*}{1.024\textwidth}{|c||c|c|c|c|c|c|c|c|}
\hline Sol & q & k & \(\alpha\) & \(\beta\) & \(\gamma\) & \(a_1\) & \(\frac{a_1}{a_2}\) & \(\frac{a_1}{a_3}\) \\
\hline  1 & 1 & 100 & 70 000 & 100 000 & 233 333 & $8.46360\times10^{-2}$ & $2.10490$  & $4.68298$  \\
\hline  2 & 1 & 100 & 70 000 & 100 000 & 233 333,3 & $8.46360\times10^{-3}$ & $2.10490$  & $4.68298$\\
\hline  3 & 1 & 100 & 70 000 & 100 000 & 233 333,33 & $8.46361\times10^{-4}$ & $2.10489$  & $4.68298$\\
\hline  4 & 1 & 100 & 70 000 & 100 000 & 233 333,333 & $8.46360\times10^{-5}$ & $2.10490$  & $4.68298$\\
\hline  5 & 1 & 100 & 70 000 & 100 000 & 233 333,33333 & $8.46360\times10^{-7}$ & $2.10490$  & $4.68298$\\
\hline
\end{tabular*}
\end{changemargin}

\bigbreak

\begin{figure}
\centering
\includegraphics[scale=0.3]{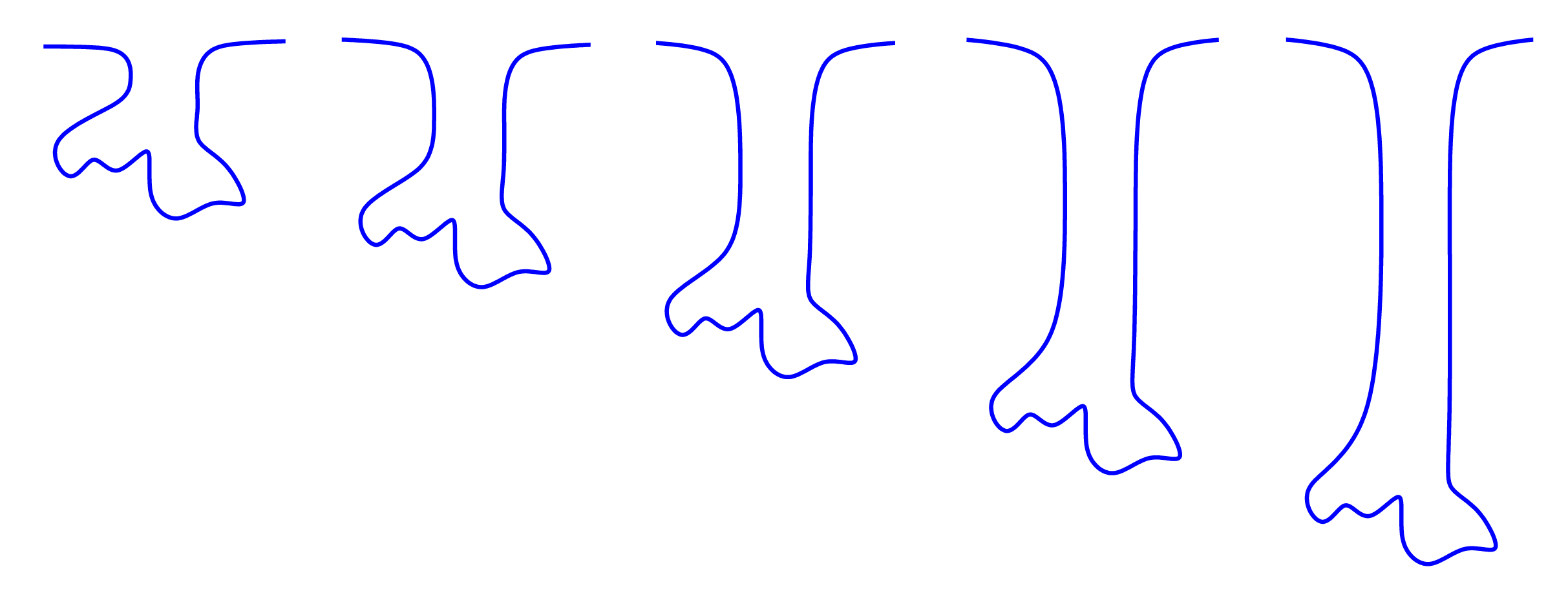}
\caption{A schematic description of a scaling geometry. The physical size of the bubbles remains the same whereas the throat of the solutions deepens \cite{Bena:2006kb}.}
\label{ScalingProcess}
\end{figure}

Therefore, when \(\gamma\) is converging towards a certain value, the supertube positions shrink drastically. When the centers get closer and closer, the size of the throat gets larger and larger as explained in \cite{Bena:2006kb} and depicted in Fig.\ref{ScalingProcess}. Because a BPS black hole has an infinite throat, the scaling process makes the bubbling solution gets more and more similar to a BPS black hole. We find a specific example of scaling microstate geometry of three BPS supertubes in \(\mathbb{R}^4\) corresponding to a BMPV microstate with $\mathcal{H} \:=\: 0.87$.

\subsubsection{Solutions with \(h \,\ne\, 0\)}
\label{subsubsec:casehzero}
Adding a non-zero constant term in \(V\) will make the Gibbons-Hawking metric look like \(\mathbb{R}^3 \times S^1\) at infinity. Such a configuration will be convenient when spectral flows and gauge transformations will be applied to reach general BPS solutions with four GH centers.
To cope with the constant term in \(V\), constant terms in \(K^I\) must be turned on to satisfy the absence of the CTCs at infinity \eqref{CTCcondition}. We fix the constant terms as follows: 
\begin{equation}
\begin{split}
& \alpha_1 \:=\: -2 h m_\infty \\
& \alpha_2 \:=\: \alpha_3 \:=\: 0.
\end{split}
\end{equation}

Indeed, such a choice ensures that

\begin{equation}
\begin{split}
&\mu\left(\infty\right) \:=\: 0 \\
&Z_I\left(\infty\right) \:=\: 1.
\end{split}
\end{equation}

We choose the following values of charges and dipole charges as an illustration:
\begin{equation}
\begin{split}
& h \:=\: 1 \\
& q \:=\: 1 \\
& k_1 \:=\: k_3 \:=\: -\:k_2 \:=\: k \:=\: 100\\
& Q_1^{(2)} \:=\: Q_3^{(2)} \:=\: \alpha \:=\: 70\,000  \\ 
& Q_3^{(1)} \:=\: Q_2^{(3)} \:=\: \beta \:=\: 100\,000  \\
& Q_2^{(1)} \:=\: Q_1^{(3)} \:=\: \gamma \:=\: 240\,000.
\end{split}
\label{SystemOfCharges2}
\end{equation} 
The bubble equations \eqref{BubbleEquation} are solvable, the no-CTC condition is satisfied and the three-center positions are
\begin{equation}
a_1 \:=\: 3.04216\ldots \times 10^{-2} \,, \qquad a_2 \:=\: 1.50188\ldots \times 10^{-2}\,, \qquad a_3 \:=\: 6.69370\ldots \times 10^{-3}.
\end{equation}

\begin{itemize}
    \item \textbf{Charges, angular momentum and entropy of the solution}
\end{itemize}

The charges and angular momentum of the solution are :

\begin{equation}
 \begin{split}
     & Q_1 \:=\: Q_3 \:=\: 75\;000\\
     & Q_2 \:=\: 45\;000\\
     & J \:=\: 5\;687\;500.\\
 \end{split}
\end{equation}

The Bekenstein-Hawking entropy of the corresponding black hole solution is

\begin{equation}
 S \:=\: 2\pi \sqrt{Q_1 Q_2 Q_3 \;-\; J^2} \:=\: 9.33592\ldots \times 10^{7}.
\end{equation}

Finally, the entropy parameter \(\mathcal{H}\)  is

\begin{equation}
 \mathcal{H} \:=\: 0.87\ldots.
\end{equation}

\begin{itemize}
    \item \textbf{Scaling solutions}
\end{itemize}

As the previous section, we scale the solution by adjusting only the value of \(\gamma\). The scaling process is summed up in the following table:

\bigbreak
\begin{center}
\begin{tabular*}{0.709\textwidth}{|c||c|c|c|c|}
\hline Solutions & \(\gamma\) & \(a_1\) & \(\frac{a_1}{a_2}\) & \(\frac{a_1}{a_3}\) \\
\hline  1 & 240 000 & $3.04216 \times 10^{-2}$ & $2.02557$  & $4.54481$  \\
\hline  2 & 234 000 & $2.97358\times10^{-3}$ & $2.09593$  & $4.66672$\\
\hline  3 & 233 400 & $2.96688\times10^{-4}$ & $2.10405$  & $4.68150$\\
\hline  4 & 233 340 & $2.96621\times10^{-5}$ & $2.10481$  & $4.68283$\\
\hline  5 &233 333,4 & $2.96614\times10^{-7}$ & $2.10490$  & $4.68299$\\
\hline
\end{tabular*}
\end{center}
\bigbreak

We find an example of scaling microstate geometry of three BPS supertubes with one Gibbons-Hawking center in Taub-NUT. Furthermore, if we compare both tables with $h=1$ and with $h=0$, the solutions are really similar to each other in the scaling limit. So the constant terms of the harmonic functions are irrelevant in the scaling limit.


\section{Constructing BPS solutions with four Gibbons-Hawking centers}


\subsection{Four-GH center bubbled solutions}

In this chapter, we consider BPS bubbled solutions with four GH centers. Those solutions can be entirely generated from the family of three-supertube solutions by using three generalized spectral flows and three gauge transformations. The eleven-dimensional metric is the same as \eqref{11dmetric}. However, we consider a much more general set of harmonic functions \(\{V,K^I,L_I,M\}\). Each function has poles at each GH center. We use the following usual notations: 
\begin{equation}
\begin{split}
    & V \:=\: q_{\infty} \:+\: \sum_{j=0}^3 \frac{q_j}{r_j}, \qquad M \:=\: m_{\infty} \:+\: \sum_{j=0}^3 \frac{m_j}{r_j} \\
    & K^I \:=\: k_{\infty}^I \:+\: \sum_{j=0}^3 \frac{k_j^I}{r_j}, \qquad L_I \:=\: l_{\infty}^I \:+\: \sum_{j=0}^3 \frac{l_j^I}{r_j},
\end{split}
\end{equation}
\(r_0\), \(r_1\), \(r_2\) and \(r_3\) are the same distances as in the first section with \(r_0\) corresponding to \(r\). In this framework, the regularity constraints are
\begin{equation}
\begin{split}
& l_j^I \:=\: -\frac{1}{2}C_{IJK} \frac{k_j^J k_j^K}{q_j} \qquad j \in \{0,\ldots,3\} \\
& m_j \:=\: \frac{1}{12}C_{IJK} \frac{k_j^I k_j^J k_j^K}{q_j^2}  \qquad j \in \{0,\ldots,3\}.
\label{constraintcharges}
\end{split}
\end{equation}

The equations \eqref{Zexpression} for the warp factors \(Z_I\), \eqref{muexpression} for \(\mu\), \eqref{omegaexpression} for \(\omega\), the bubble equations \eqref{BuEQ} and the no-CTC condition \eqref{CTCcondition} are still valid.


\subsection{Spectral flows and gauge transformations}

For BPS solutions, generalized spectral flows and gauge transformations are well-understood \cite{Bena:2008wt,DallAgata:2010srl}. They mix the harmonic functions leaving the bubble equations and the entropy function invariant. Furthermore, they send the family of three-supertube solutions to the family of solutions with four GH centers. The harmonic functions \(\{ V,K^I,L_i,M \}\) of BPS solutions transform under three gauge transformations parameterized by three constants \(g^I\) as \cite{Bena:2005ni}:
\begin{equation}
\begin{aligned}
&V \:\rightarrow\: V, \qquad K^{I} \:\rightarrow\: K^{I} \:+\: g^I V, \\
&L^{I} \:\rightarrow\: L^{I} \:-\: C_{IJK}\, g^J K^K \:-\: \frac{1}{2}  C_{IJK}\, g^J g^K V, \\
&M \:\rightarrow\: M \:-\: \frac{1}{2} g^I L^I \:+\: \frac{1}{4} C_{IJK}\, g^I g^J K^K \:+\: \frac{1}{12} C_{IJK} \,g^I g^J g^K V,
\label{gaugetransformation}
\end{aligned}
\end{equation}

and under a family of generalized spectral flows parameterized by three constants \(\gamma^I\) as \cite{Bena:2008wt}:
\begin{equation}
\begin{aligned}
&M \:\rightarrow\: M, \qquad L^{I} \:\rightarrow\: L^{I} \:-\: 2\, \gamma^I M, \\
&K^{I} \:\rightarrow\: K^{I} \:-\: C_{IJK}\, \gamma^J L^K \:+\:  C_{IJK}\, \gamma^J \gamma^K M, \\
&V \:\rightarrow\: V \:+\:  \gamma^I K^I \:-\: \frac{1}{2} C_{IJK}\, \gamma^I \gamma^J L^K \:+\: \frac{1}{3} C_{IJK} \,\gamma^I \gamma^J \gamma^K M.
\label{spectralflow}
\end{aligned}
\end{equation}


\subsection{One particular example of solutions}
\label{subsec:process4GH}

We start from the particular solution of three supertubes in Taub-NUT background of section \ref{subsubsec:casehzero} with \(h \,\ne\, 0\) and \(\gamma \,=\, 233\,400\). We perform three independent spectral flows then three independent gauge transformations. To obtain physical solutions and a Gibbons-Hawking metric which looks like \(\mathbb{R}^4\) at infinity, the transformations are constrained such that in the resulting solution:
\begin{itemize}
    \item The constant term in $V$ is 0.
    \item The sum of all charges in $V$ is 1.
    \item The charges in $V$ are integers.
    \item The constant terms in all \(K^I\) are 0.
    \item The values of the charges and dipole charges are rational \cite{Balasubramanian:2006gi}.
\end{itemize}

To satisfy the conditions above, we first leave a non-zero constant term in $V$ in order to remove the constant terms in each \(K^I\) with three gauge transformations. Then we fix one spectral flow parameter to satisfy the second condition. The third condition cannot be easily satisfied. So, we fix the two remaining spectral flow parameters to have the closest values to integers. Afterwards, all six transformation parameters are fixed. We remove by hand the constant term in $V$. Because the bubble equations are changing significantly, we have to solve them accordingly. For this purpose, two options exist :
\begin{itemize}
\item We solve the bubble equations by considering the positions of the centers as variables and all the charges are fixed. 
\item We solve the bubble equation by considering the charges as variables and the positions are fixed. In the scaling limit, one can see straightforwardly that an infinitesimal change of three dipole charges ($k_0^1$, $k_0^2$ and $k_0^3$ for instance) can annihilate a change of order one on the right hand side of the bubble equations \eqref{BuEQ}. Consequently, one can find a solution to the bubble equations only perturbing infinitesimally the values of the charges which ensures that no closed timelike curves will appear. 
\end{itemize}
In general, we apply the first method. However, when it was not possible to do, we apply the second one. \\
At that point, the values of charges and dipole charges are real numbers with an infinite number of digits. The last step is to round the values taking into account the necessary regularity conditions \eqref{constraintcharges}. Once again, the solutions are not invariant, we have to check that the bubble equations are still solvable and that the no-CTC condition is still satisfied. \\
We present the result of applying the above procedures on the three-supertube configuration with \(h \,\ne\, 0\) and \(\gamma \,=\, 233\,400\):

\begin{equation}
\begin{split}
& V \:=\: \frac{1}{r} \:+\: \frac{1}{r_1} \:+\: \frac{12}{r_2} \:-\: \frac{13}{r_3} \\
& K^1 \:=\:  -\frac{2087}{10000}\frac{1}{r} \:-\: \frac{678089}{1250}\frac{1}{r_1} \:+\: \frac{55636379}{10000}\frac{1}{r_2} \:+\: \frac{3445309}{2000}\frac{1}{r_3} \qquad \qquad \qquad \qquad \qquad \qquad \qquad \qquad \\
& K^2 \:=\:  -\frac{491}{2500}\frac{1}{r} \:+\: \frac{4712993}{1250}\frac{1}{r_1} \:+\: \frac{30306499}{5000}\frac{1}{r_2} \:+\: \frac{32175101}{5000}\frac{1}{r_3} \nonumber \\
\end{split}
\end{equation} 
\begin{equation}
\begin{split}
& K^3 \:=\:  \frac{1}{10000}\frac{1}{r} \:-\: \frac{49939}{10000}\frac{1}{r_1} \:-\: \frac{311181}{5000}\frac{1}{r_2} \:+\: \frac{133657}{2000}\frac{1}{r_3}  \\
& L^1 \:=\:  1 \:+\: \frac{491}{25000000}\frac{1}{r} \:+\: \frac{235362157427}{12500000}\frac{1}{r_1} \:+\: \frac{3143602221773}{100000000}\frac{1}{r_2} \:+\: \frac{4300427474357}{130000000}\frac{1}{r_3}  \\
& L^2 \:=\: 1 \:+\: \frac{2087}{100000000}\frac{1}{r} \:-\: \frac{33863086571}{12500000}\frac{1}{r_1} \:+\: \frac{5770994684533}{200000000}\frac{1}{r_2} \:+\: \frac{460489665013}{52000000}\frac{1}{r_3}  \\
& L^3 \:=\:  1 \:-\: \frac{1024717}{25000000}\frac{1}{r} \:+\: \frac{3195828710377}{1562500}\frac{1}{r_1} \:-\: \frac{1686143864527121}{600000000}\frac{1}{r_2} \:+\: \frac{110853165051209}{130000000}\frac{1}{r_3}  \\
& M \:=\: -\frac{115048645}{10000} \:-\: \frac{1024717}{500000000000}\frac{1}{r} \:+\: \frac{159596489967517003}{31250000000}\frac{1}{r_1} \\
& \qquad \:-\: \frac{174898644635804679967}{24000000000000}\frac{1}{r_2} \:+\: \frac{14816301481249441313}{6760000000000}\frac{1}{r_3}.
\label{NewSol}
\end{split}
\end{equation} 

The bubble equations are solved by the following positions:
\begin{equation}
 a_1 \:=\:2.67495\ldots\times10^{-2} \,, \qquad a_2 \:=\: 1.27076\ldots\times10^{-2} \,, \qquad a_3 \:=\:5.70977\ldots\times10^{-3}.
\end{equation}

\begin{itemize}
    \item \textbf{Charges, angular momentum and entropy}
\end{itemize}

The charges and angular momentum of the solution are:

\begin{equation}
 \begin{split}
     & Q_1 \:=\: \frac{99858458954459}{1300000000}\\
     & Q_2 \:=\: \frac{83964235108323}{2600000000}\\
     & Q_3 \:=\: \frac{856306630373655247}{7800000000}\\
     & J \:=\: \frac{705272590995929898902049}{1352000000000000}.
 \end{split}
\end{equation}

The Bekenstein-Hawking entropy of the corresponding black hole solution is

\begin{equation}
 S \:=\: 9.12620\ldots \times 10^{7}.
\end{equation}

Finally, the entropy parameter \(\mathcal{H}\)  is

\begin{equation}
 \mathcal{H} \:=\: 0.000775.
\end{equation}

The angular momentum is extremely close to the maximally spinning value. This is not a coincidence and this is one of the key points of our paper. Picking random three-supertube configuration also gives a near-maximally spinning solution after spectral flows and gauge transformations. In order to prevent this, a very fine tuning between the parameters of the three-supertube solution is required. The main reason is that the charges \(Q_1\), \(Q_2\) and \(Q_3\) strongly increase under spectral flows if all the conditions imposed for $V$ are satisfied. Consequently, the angular momentum is forced to reach a value close to \(\sqrt{Q_1 Q_2 Q_3}\) to leave the entropy function invariant. This will be fully described in section \ref{subsec:nonmax}.


\subsubsection{Scaling solutions}
\label{subsubsec:scalsol}
We check whether the new solution can still scale. We only change \(k_2^1\), but any other $K^I$ coefficient would have worked. Furthermore, we take into account that any change of \(k_2^1\) produces a change in the \(L^I\)'s charges and in M according to \eqref{constraintcharges}. We give the steps in the following table: 

\bigbreak
\begin{center}
\begin{tabular*}{0.693\textwidth}{|c||c|c|c|c|}
\hline Sol & $k_2^1$ & \(a_1\) & \(\frac{a_1}{a_2}\) & \(\frac{a_1}{a_3}\) \\
\hline  1 & 5563.6379  & $2.67494\times10^{-2}$ & $2.10499$  & $4.68485$  \\
\hline  2 & 5562.9979  & $5.47309\times10^{-3}$ & $2.10489$  & $4.68358$\\
\hline  3 & 5562.8379 & $1.52860\times10^{-4}$ & $2.10487$  & $4.68327$\\
\hline  4 & 5562.83334 & $1.22654\times10^{-6}$ & $2.10486$  & $4.68326$\\
\hline  5 & 5562.822208 & $1.62447\times10^{-7}$ & $2.10487$  & $4.68325$\\
\hline  6 & 5562.8333032 & $2.83275\times10^{-9}$ & $2.10487$  & $4.68325$\\
\hline  7 & 5562.8333031  & $1.72517\times10^{-10}$ & $2.10486$  & $4.68324$\\
\hline
\end{tabular*}
\end{center}
\bigbreak

Thus, we built asymptotically \(\mathbb{R}^4\) solutions with a shrinking cluster of four bubbled GH centers with no difference in scales. These are microstates of a black hole with a macroscopically large horizon. In the next section we investigate how systematic the large $J$ is.

\section{Discussion}
\label{subsec:nonmax}

\subsection{Solutions with four GH centers with no difference in scales are necessarily near-maximally spinning}

Understanding the impact of spectral flows on a cluster of three supertubes in Taub-NUT by an analytical approach is complicated due to the number of parameters and the complexity of the equations. However, by a numerical approach, it appears that performing three constrained spectral flows on a three-supertube solution with no difference in scales between the inter-center distances (that is to say \(a_3\), \(a_2-a_3\) and \(a_1-a_2\) are of the same order of magnitude) will result in an entropy parameter extremely close to 0. By ``constrained spectral flows'' one means that the resulting harmonic function $V$ must have integer charges whose sum is equal to one.  We reached this conclusion after having studied many different examples which have led to a value of \(\mathcal{H}\) around 0.01 each time. Insights of our approach are given in the next two sections and in appendix \ref{sec:numerics}. 

\subsubsection{Numerical analysis}
\label{subsubsec:NumAnal}
First, we investigate all the solutions with four GH centers where the centers scale as follows:
\begin{equation}
\begin{split}
\frac{a_1 - a_2}{a_3} \:\approx\: 1 \\
\frac{a_2 - a_3}{a_3} \:\approx\: 1.
\end{split}
\label{distanceratios}
\end{equation}
We would like to scan all the solutions of this kind, to compute their entropy parameters, and to find out if they are indeed all of the order of 0.01. \\
Let us list all the free parameters. The starting configuration is a three-supertube solution with six positive charges $Q_I^{(J)}$, three dipole charges $k_I$ and one Gibbons-Hawking charge $q$. The constant terms do not enter in the formula for the entropy parameter and hence are irrelevant. The scaling conditions fix three charges. Furthermore, we notice that changes of $q$ can be reabsorbed into changes of dipole charges. So without loss of generality we can consider that $q$ is equal to one. We can also restrict the sign changes of dipole charges by symmetry. Then, all six parameters of the spectral flows and the gauge transformations are constrained to cancel the constant terms in all $K^I$ and to have a Gibbons-Hawking harmonic function of the form:
 \begin{equation}
 V \:=\: \frac{1}{r} \:-\: \frac{2}{r_1} \:+\: \frac{1}{r_2} \:+\: \frac{1}{r_3}.
 \label{GHhamonicfunc}
 \end{equation}
This constitutes the best choice to maximize the value of the entropy parameter as shown at the end of appendix \ref{sec:numerics}. The process to obtain such solutions is the same as the one given in section \ref{subsec:process4GH}. Finally, one needs to vary only three charge ratios, the norms and some signs of the dipole charges to scan all the solutions satisfying \eqref{distanceratios} and \eqref{GHhamonicfunc}. We built a huge number of such four-center solutions using a loop algorithm. We have computed their entropy parameter to establish the variation of the entropy parameter as a function of the free parameters. We concluded that in any situation the entropy parameter is at most of order of 0.01. Then, we extended to solutions which do not necessarily satisfy \eqref{distanceratios} and \eqref{GHhamonicfunc}. We observed the same feature. All the details are presented in appendix \ref{sec:numerics}.

\subsubsection{An analytical argument}
We investigate the near-maximally spinning feature in the context of the specific choice of charges and dipole charges made in \ref{subsubsec:specificchoice}. With this specific choice it is easier to understand analytically why the conditions imposed to the spectral flows produce an important increase in the charges \(Q_1\), \(Q_2\) and \(Q_3\) which requires an increase in $J$ to leave the entropy invariant. We sum up how we proceed in the following steps:
\begin{itemize}
    \item We express the charges \(\alpha\), \(\beta\) and \(\gamma\) using the scaling condition relation \eqref{constraintscaling} to reduce the number of independent parameters. We obtain approximations as \(F\left(\frac{a_1}{a_3},\frac{a_2}{a_3}\right)4 k^2\) for each charge where the $F$'s are rational fractions.
    \item Assuming no difference in scales (for instance \(1.2\; a_3 < 1.1\; a_2 < a_1 < 10\; a_3 \)) allows to restrict the functions \(F\left(\frac{a_1}{a_3},\frac{a_2}{a_3}\right)\) to certain ranges of values for each charge. 
    \item We analytically express how \(Q_1\), \(Q_2\) and \(Q_3\) transform under three generalized spectral flows one of which is fixed by requiring the sum of the charges in $V$ to be equal to one.
    \item For each domain of \(F\left(\frac{a_1}{a_3},\frac{a_2}{a_3}\right)\) we show numerically that if \(Q_1\), \(Q_2\) and \(Q_3\) do not double after spectral flows, the spectral flow parameters must be in the same order of magnitude as \(\frac{k}{\alpha}\) at most.
    \item Finally, by studying the transformation of \(V\) under spectral flows \eqref{spectralflow} in the specific domains of parameters found above, this gives rise to a function $V$ with charges smaller than one which is non-physical.
\end{itemize}
With those two analyses, we conclude that any BPS bubbled scaling solution with four GH centers with no difference in scales is necessarily near-maximally spinning. \\
\newline
By the same kind of arguments, if no assumption about the positions is made, building a solution which is not a near-maximally spinning solution is also difficult but not impossible \cite{Heidmann:2017}. \\


\section{Conclusion}

We have built several examples of four-center BPS bubbled solutions. Starting with three supertubes in Taub-NUT, we gave a systematic protocol and we built several scaling solutions. Secondly, using generalized spectral flows we generated BPS solutions with four Gibbons-Hawking centers. We gave a numerical protocol to obtain such solutions from three-supertube scaling solutions and we built specific examples. We reached one important conclusion: BPS solutions with four GH centers that do not have an entropy parameter close to 0 are rare. Furthermore, when such solutions have no difference in scales between the inter-center distances, they are necessarily near-maximally spinning. \\

There are several interesting directions for future research. The first is to find BPS solutions with four GH centers which are not near-maximally spinning. We expect them to display a difference in scales between the inter-center distances, as in the examples built in \cite{Bena:2006kb}. Using algorithms to fine-tune the initial three-supertube parameters, one can plan to find such solutions \cite{Heidmann:2017}. One can also investigate different four-center constructions with one supertube and three Gibbons-Hawking centers. Those solutions are smooth in the D1-D5-P duality frame and can be generated from three-supertube solutions by two spectral flows. Thus, one may expect them not to necessarily be near-maximally spinning solutions even when the centers have no difference in scale. Secondly, one can extend our work to almost-BPS four-center solutions \cite{Goldstein:2008fq} following the same ideas. Thirdly one can add non-Abelian hair to the solutions as it was done in \cite{Ramirez:2016tqc}. Finally, one can also investigate whether solutions with more than four Gibbons-Hawking centers without difference in scale are also near-maximally spinning and, more generally, how universal is the feature we found.

\section*{Acknowledgements}

I would like to thank Iosif Bena and Pedro Ramirez for the interesting discussions.  This work was supported by an ENS Lyon doctoral grant. 

\appendix
\section{Numerical analysis of the entropy parameter of four-GH center solutions}
\label{sec:numerics}

\begin{figure}
 \noindent\hfil\rule{0.7\textwidth}{.6pt}\hfil \\
 \centering
\begin{tabular}{ccc}
\addlinespace[1ex]
\subf{\includegraphics[width=70mm]{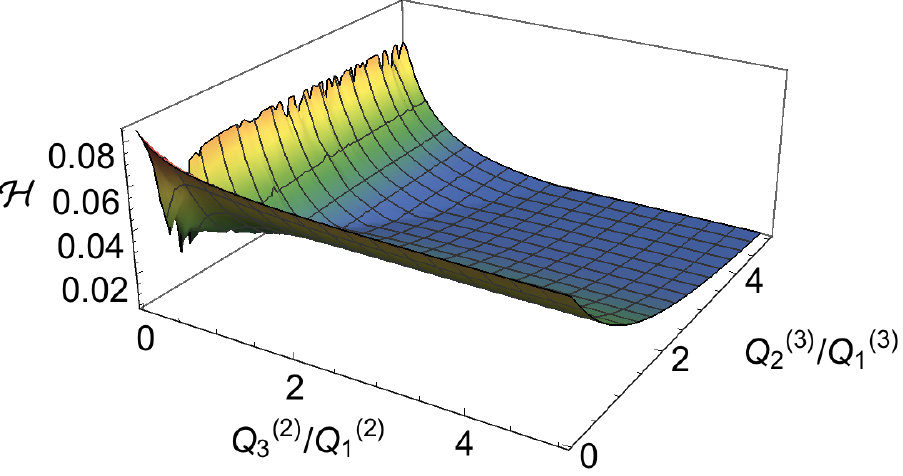}}
     {\\
     (a) $\frac{Q_2^{(1)}}{Q_1^{(3)}} \:=\: 0.1$}
&
\subf{\includegraphics[width=70mm]{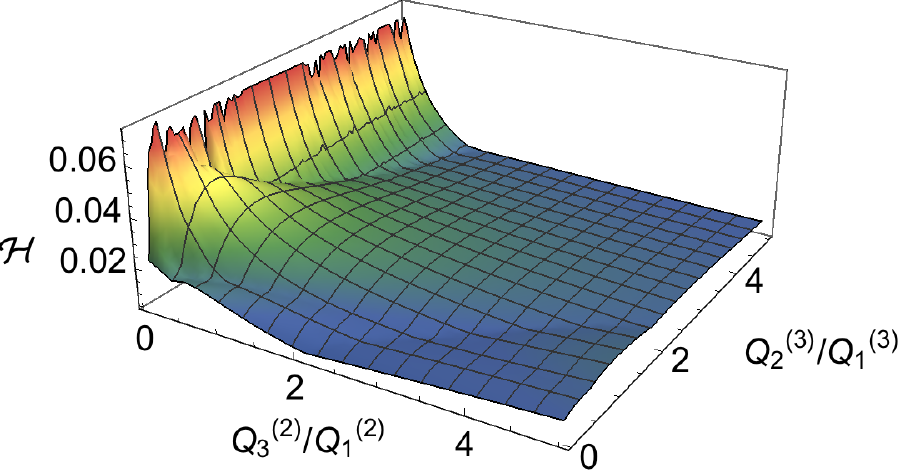}}
     {\\
     (b) $\frac{Q_2^{(1)}}{Q_1^{(3)}} \:=\: 0.5$}
\\
 \addlinespace[2ex]
\subf{\includegraphics[width=70mm]{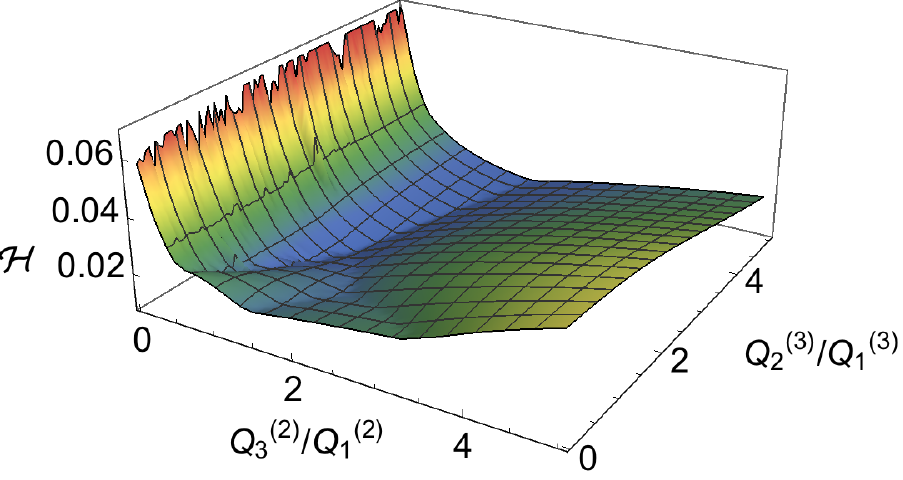}}
     {\\
     (c) $\frac{Q_2^{(1)}}{Q_1^{(3)}} \:=\: 1$}
&
\subf{\includegraphics[width=70mm]{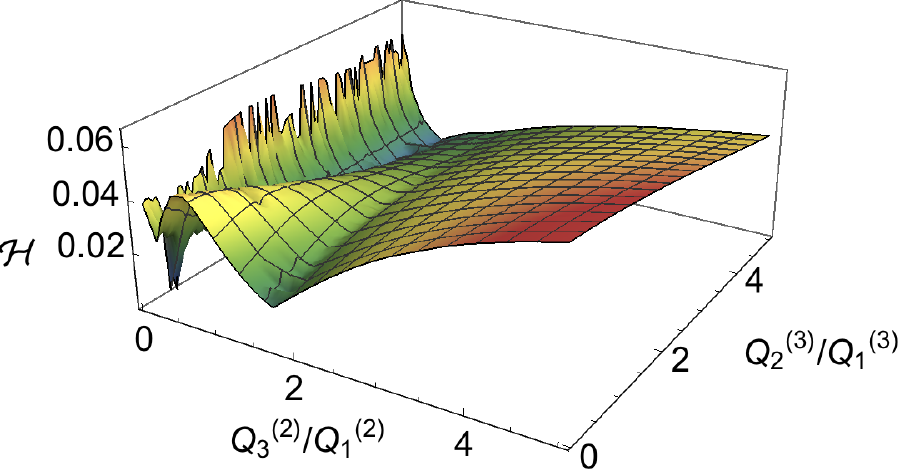}}
     {\\
     (d) $\frac{Q_2^{(1)}}{Q_1^{(3)}} \:=\: 2$}
\\
 \addlinespace[2ex]
\subf{\includegraphics[width=70mm]{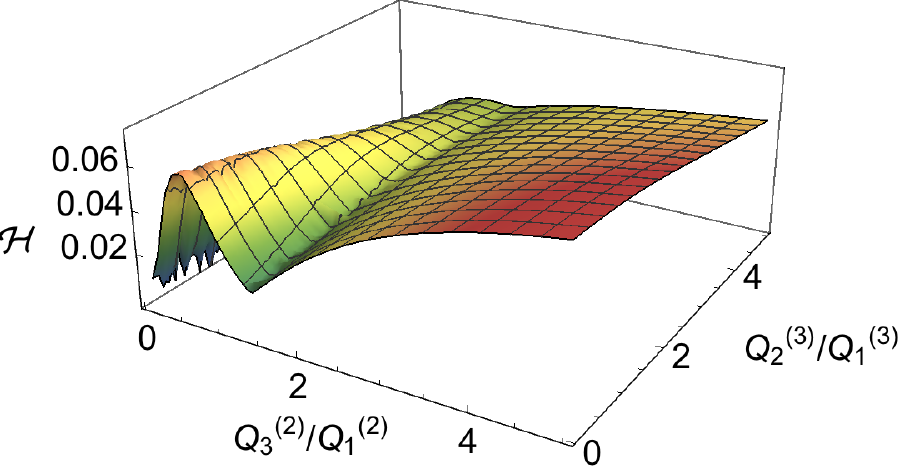}}
     {\\
     (e) $\frac{Q_2^{(1)}}{Q_1^{(3)}} \:=\: 5$}
&
\subf{\includegraphics[width=70mm]{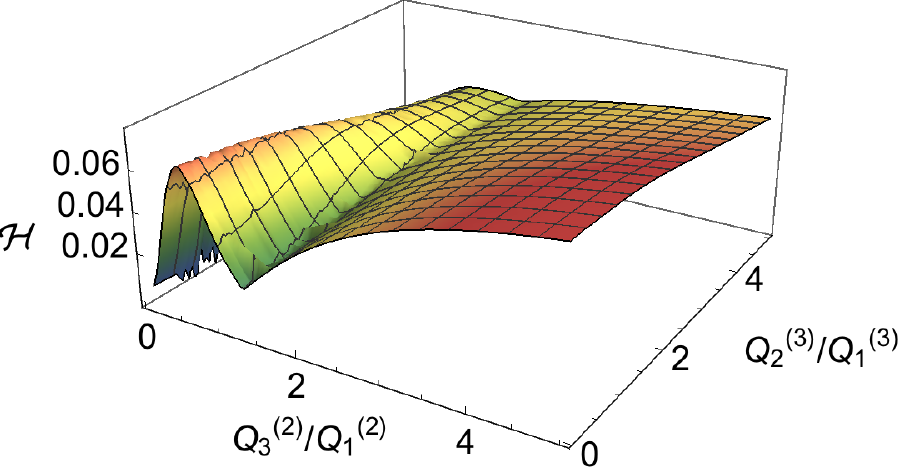}}
     {\\
     (f) $\frac{Q_2^{(1)}}{Q_1^{(3)}} \:=\: 10$}
\\
\addlinespace[1ex]
\end{tabular}
\noindent\hfil\rule{0.7\textwidth}{.6pt}\hfil
\caption{The entropy parameter $\mathcal{H}$ as a function of the charge ratios with $q$, $k_1$, $-k_2$ and $k_3$ equal to 1.}
\label{entropyparamchargeratios}
\end{figure}

In this section, we detail how we scanned all the different values of the entropy parameter we can get from scaling solutions with four Gibbons-Hawking centers. As explained in section \ref{subsubsec:NumAnal}, we first scanned all the scaling solutions which satisfy
\begin{equation}
\begin{split}
& \frac{a_1 - a_2}{a_3} \:\approx\: 1 \\
& \frac{a_2 - a_3}{a_3} \:\approx\: 1 \\
& V \:=\: \frac{1}{r} \:-\: \frac{2}{r_1} \:+\: \frac{1}{r_2} \:+\: \frac{1}{r_3},
\end{split}
\end{equation}
by varying the free parameters of the constructions: three charge ratios ($\frac{Q_2^{(1)}}{Q_1^{(3)}}$, $\frac{Q_3^{(2)}}{Q_1^{(2)}}$ and $\frac{Q_2^{(3)}}{Q_3^{(1)}}$ for instance) and the three dipole charges $k_1$, $k_2$ and $k_3$.
The graphs Fig.\ref{entropyparamchargeratios} give the variations of the entropy parameter according to three charge ratios from 0.1 to 4 with $q$, $k_1$, $-k_2$ and $k_3$ equal to 1. Each graph has been built with 10000 scaling solutions which are not necessarily CTC-free and the result has been interpolated and smoothed for readability. Thus, it is important to point out that a point of one curve does not necessarily correspond to a CTC-free solution with four GH centers. Nevertheless, any CTC-free four-center solution meeting the conditions above has an entropy parameter near or below the curve's prediction.

\begin{figure}
\centering
\noindent\hfil\rule{0.7\textwidth}{.6pt}\hfil \\
\begin{changemargin}{-0.5cm}{-0.5cm}
\begin{tabular}{cc}
\addlinespace[2ex]
\subf{\includegraphics[width=70mm]{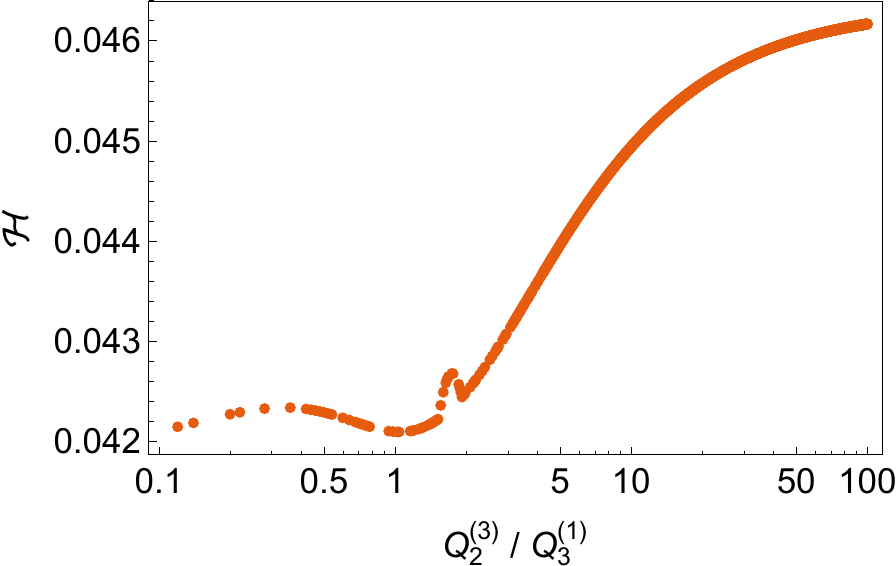}}
     {\\
     (a) $\frac{Q_2^{(1)}}{Q_1^{(3)}} \:=\: 1$ and $\frac{Q_3^{(2)}}{Q_1^{(2)}} \:=\: 0.2$}
& \hspace{1cm}
\subf{\includegraphics[width=70mm]{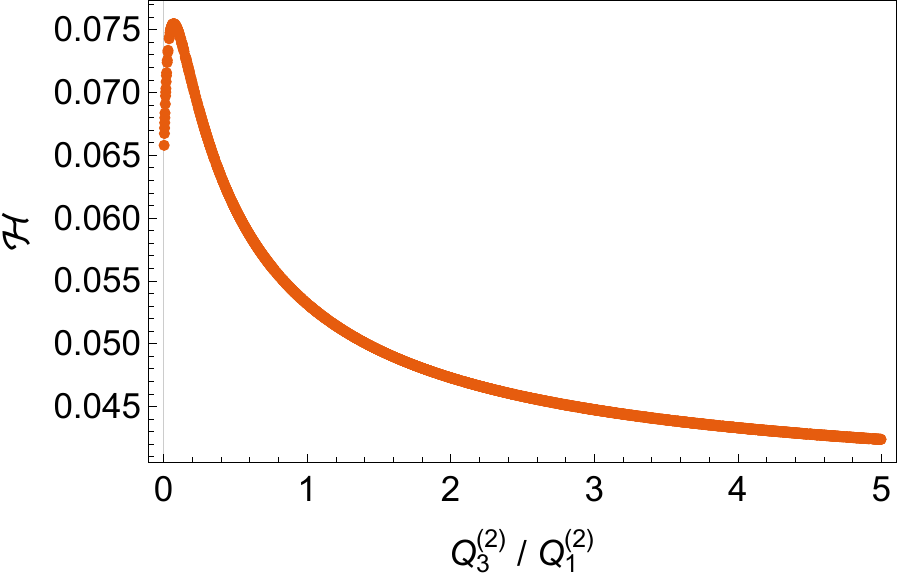}}
     {\\
     (b) $\frac{Q_2^{(1)}}{Q_1^{(3)}} \:=\: 0.2$ and $\frac{Q_2^{(3)}}{Q_3^{(1)}} \:=\: 0.2$}
\\
 \addlinespace[2ex]
\subf{\includegraphics[width=70mm]{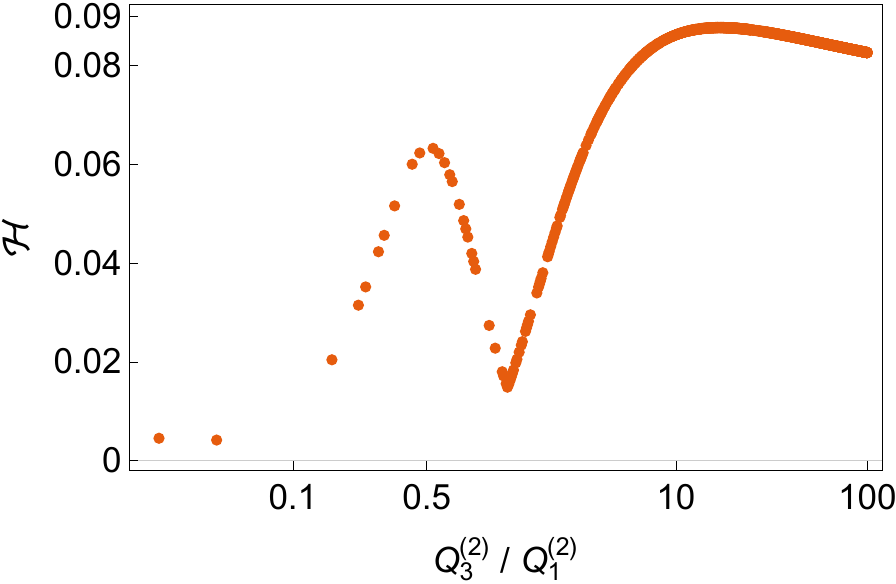}}
     {\\
     (c)  $\frac{Q_2^{(1)}}{Q_1^{(3)}} \:=\: 7$ and $\frac{Q_2^{(3)}}{Q_3^{(1)}} \:=\: 0.003$}
&
\subf{\includegraphics[width=70mm]{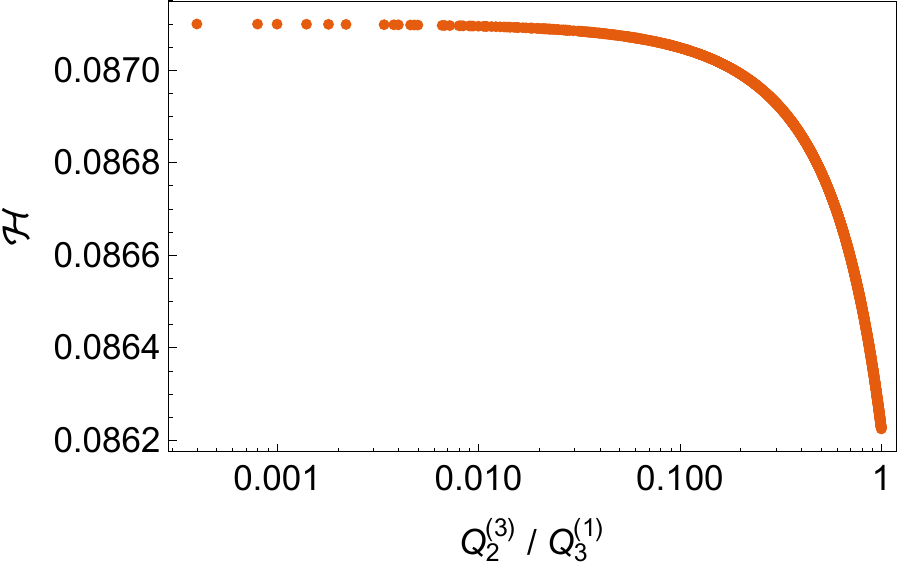}}
     {\\
     (d) $\frac{Q_2^{(1)}}{Q_1^{(3)}} \:=\: 7$ and $\frac{Q_3^{(2)}}{Q_1^{(2)}} \:=\: 12$}
\\
\addlinespace[2ex]
\end{tabular}
\end{changemargin}
\noindent\hfil\rule{0.7\textwidth}{.6pt}\hfil
\caption{The entropy parameter $\mathcal{H}$ as a function of one of the charge ratios with  $q$, $k_1$, $-k_2$ and $k_3$ equal to 1.}
\label{entropyparamchargeratiosslice}
\end{figure}

In the domains studied, the entropy parameter is indeed of order 0.01. One also has to check the regions where $\mathcal{H}$ increases. For this purpose, we have investigated the evolution of the entropy parameter as a function of one of the charge ratios for a larger range of values. We observed that the evolutions for any value of the two other charge ratios are similar to the four evolutions shown in Fig.\ref{entropyparamchargeratiosslice} and that $\mathcal{H}$ is bounded between 0 and 0.09. However it is important to point out that in the domains of very large or very small charge ratios the solutions can possibly have closed timelike curves. Indeed, if we want the four centers to scale in one block, the bubble equations \eqref{BuEQ} imply that one supertube at least must have a negative dipole charge. Consequently, if some charges are small compared to the value of the dipole charge, the warp factor \eqref{Zexpression} of the corresponding supertube may be negative around other poles. Thus, the upper bound 0.09 of the entropy parameter can be thought of as a generous upper bound for CTC-free solutions.

\begin{figure}
\centering
\includegraphics[width=100mm]{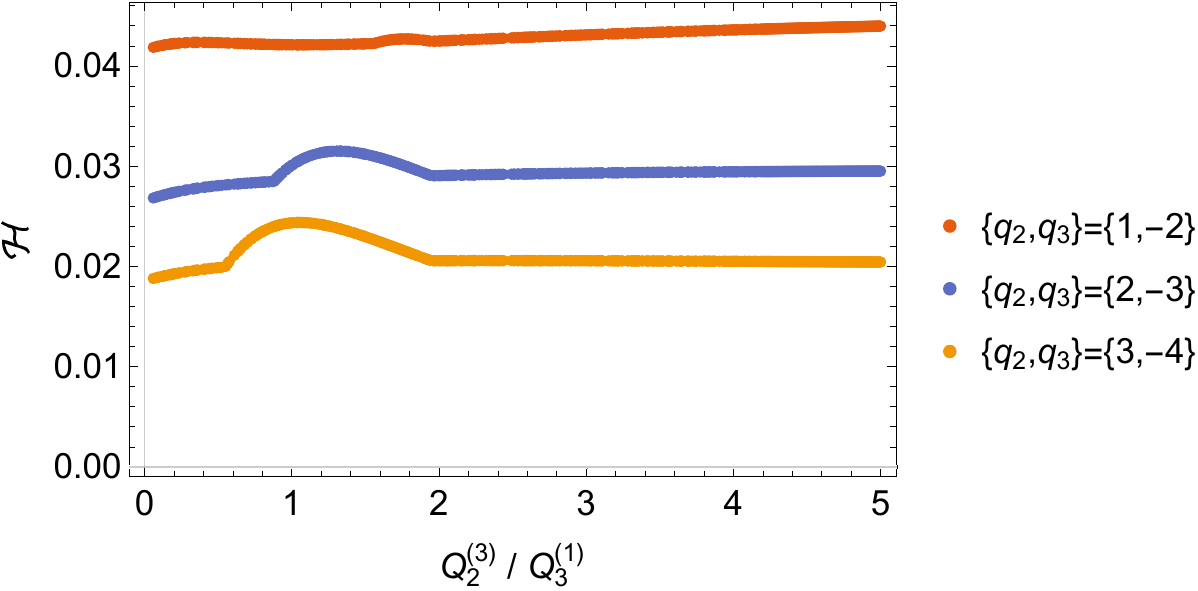}
\caption{The entropy parameter $\mathcal{H}$ as a function of one of the charge ratios for three different GH functions $V$ with $q \:=\: 1$, $k_1\:=\: -k_2\:=\: k_3 \:=\: 1$, $\frac{Q_2^{(1)}}{Q_1^{(3)}} \:=\: 1$ and $\frac{Q_3^{(2)}}{Q_1^{(2)}} \:=\: 0.2$.}
\label{entropyparamVcharge}
\end{figure}

\begin{figure}
\noindent\hfil\rule{0.7\textwidth}{.6pt}\hfil \\
\centering
\begin{tabular}{cc}
\addlinespace[1ex]
\subf{\includegraphics[width=70mm]{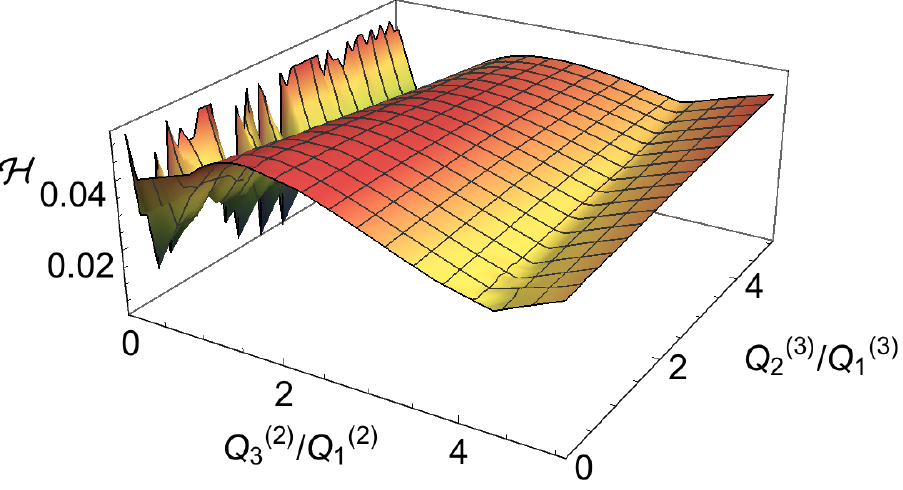}}
     {\\
     (a) $\frac{a_1}{a_2} \:\approx\: 1.1$}
&
\subf{\includegraphics[width=70mm]{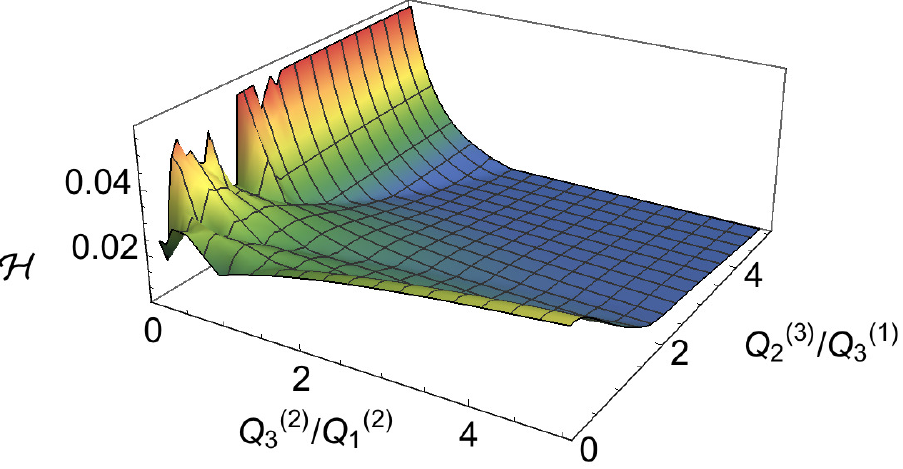}}
     {\\
     (b) $\frac{a_1}{a_2}  \:\approx\: 3.1$}
\\
 \addlinespace[2ex]
\subf{\includegraphics[width=70mm]{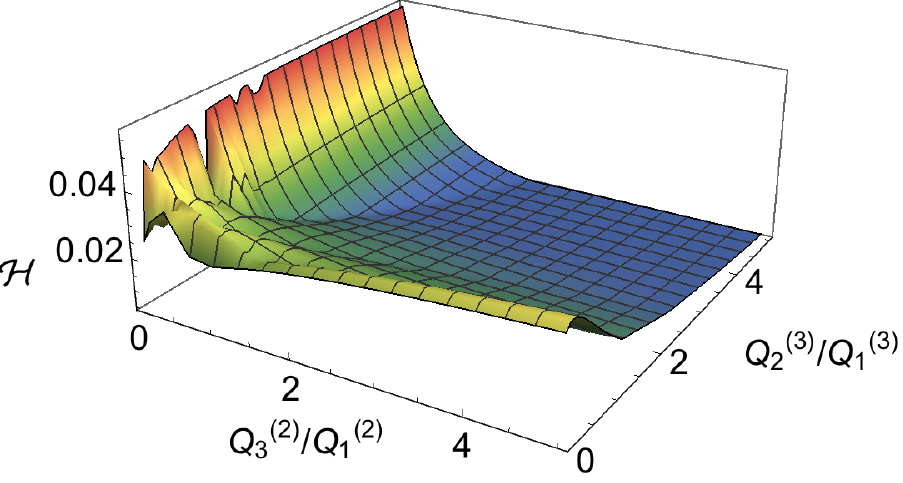}}
     {\\
     (c) $\frac{a_1}{a_2}  \:\approx\: 5.6$}
&
\subf{\includegraphics[width=70mm]{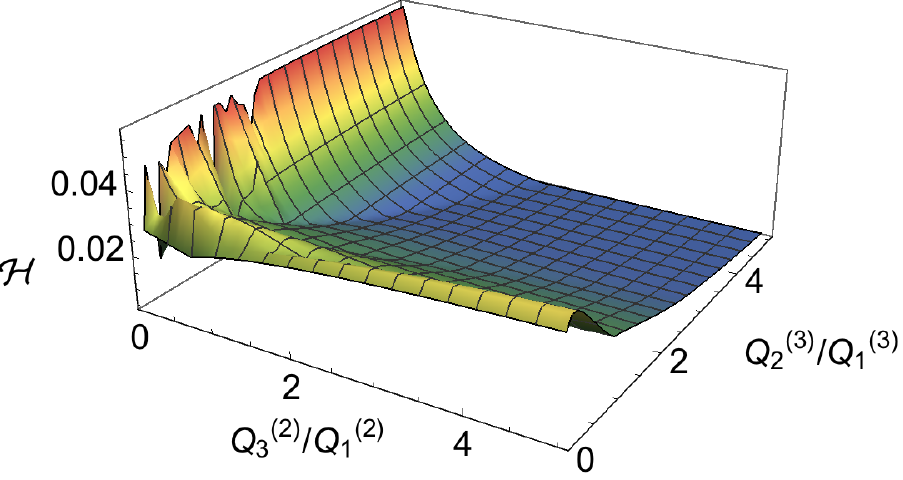}}
     {\\
     (d) $\frac{a_1}{a_2}  \:\approx\: 10.1$}
\\
\addlinespace[1ex]
\end{tabular}
\noindent\hfil\rule{0.7\textwidth}{.6pt}\hfil
\caption{The entropy parameter $\mathcal{H}$ as a function of two charge ratios and one distance ratio with  $q$, $k_1$, $-k_2$ and $k_3$ equal to 1.}
\label{entropyparamdistratios}
\end{figure}

Similarily, we studied the entropy parameter as a function of the dipole charges and their signs. We found in all configurations the same kind of evolutions as in Fig.\ref{entropyparamchargeratios}. Consequently, any entropy parameter of solutions with four Gibbons-Hawking centers satisfying \eqref{distanceratios} and \eqref{GHhamonicfunc} are of the order of 0.01.
The last step is to ensure that solutions which do not satisfy \eqref{distanceratios} and \eqref{GHhamonicfunc} have the same property. The graph Fig.\ref{entropyparamVcharge} illustrates that if the GH function $V$ has higher charges the entropy parameter is getting closer to 0. So the condition \eqref{GHhamonicfunc} gives the solutions which have the highest values of $\mathcal{H}$. We vary also the condition \eqref{distanceratios} taking into account that the inter-center distances must remain of the same order. The graphs Fig.\ref{entropyparamdistratios} show that changing the aspect ratio $\frac{a_1}{a_2}$ does not produce a significant change of the entropy parameter. \\
 
This exhaustive numerical analysis shows that any solution with four GH centers with no difference in scales is necessarily near-maximally spinning.

\newpage

\bibliography{refs}
\bibliographystyle{utphysmodb}

\end{document}